\documentclass[iop,twocolumn]{aastex6}
\usepackage{amsmath}
\usepackage{url}
\usepackage{enumerate}
\usepackage{multirow}

\shorttitle{Intermittent pulsars}
\shortauthors{Wang et al.}

\begin{document}
\title{Radio observations of two intermittent pulsars:  PSRs J1832+0029 and J1841$-$0500}
\email{wangshuangqiang@xao.ac.cn}
\author{S. Q. Wang,\altaffilmark{1,2,3}, J. B. Wang,\altaffilmark{1,4,5}, G. Hobbs\altaffilmark{3}, S.B. Zhang,\altaffilmark{2,3,6}, R.~M.~Shannon\altaffilmark{7}, S. Dai\altaffilmark{3}, R. Hollow\altaffilmark{3}, M. Kerr\altaffilmark{8}, V. Ravi\altaffilmark{9}, N. Wang\altaffilmark{1,4,5}, L. Zhang\altaffilmark{2,3,10}}
\altaffiltext{1}{Xinjiang Astronomical Observatory, Chinese Academy of Sciences, Urumqi, Xinjiang 830011, China}
\altaffiltext{2}{University of Chinese Academy of Sciences, Beijing 100049, China}
\altaffiltext{3}{CSIRO Astronomy and Space Science, Australia Telescope National Facility, PO Box 76, Epping, NSW 1710, Australia}
\altaffiltext{4}{Key Laboratory of Radio Astronomy, Chinese Academy of Science, 150 Science 1-Street, Urumqi, Xinjiang, China, 830011}
\altaffiltext{5}{Xinjiang Key Laboratory of Radio Astrophysics, 150 Science1-Street, Urumqi, Xinjiang, 830011, China}
\altaffiltext{6}{Purple Mountain Observatory, Chinese Academy of Sciences, Nanjing 210008, China}
\altaffiltext{7}{Centre for Astrophysics and Supercomputing, Swinburne University of Technology, P.O. Box 218, Hawthorn, Victoria 3122, Australia}
\altaffiltext{8}{Naval Research Laboratory, 4555 Overlook Ave., SW, Washington, DC 20375, USA}
\altaffiltext{9}{Cahill Center for Astronomy and Astrophysics, MC 249-17, California Institute of Technology, Pasadena, CA 91125, USA.}
\altaffiltext{10}{National Astronomical Observatories, Chinese Academy of Sciences, Beijing 100101, China}

\begin{abstract}
We present long-term observations of two intermittent pulsars, PSRs~J1832+0029 and J1841$-$0500 using the Parkes 64\,m radio telescope.  The radio emission for these pulsars switches ``off'' for year-long durations.  Our new observations have enabled us to improve the determination of the on-off timescales and the spin down rates during those emission states.  In general our results agree with previous studies of these pulsars, but we now have significantly longer data spans.  We have identified two unexpected signatures in the data. Weak emission was detected in a single observation of PSR~J1832$+$0029 during an ``off'' emission state. For PSR~J1841$-$0500, we identified a quasi-periodic fluctuation in the intensities of the detectable single pulses, with a modulation period between 21 and 36 pulse periods. 
 
\end{abstract}


\section{Introduction}

Radio pulsars are fast rotating and highly magnetised neutron stars which generate beams of radio emission. The  radio emission from some pulsars varies significantly on various timescales.  \citet{Backer70} reported the phenomenon of pulse nulling, in which the pulsed emission ceases suddenly and then remains ``off'' for timescales of several pulse periods to hours or even days.  Intermittent pulsars present similar behaviour, but have much longer ``on'' or ``off'' timescales which can last between several days to years~\citep{Kramer06}. 

Studies of the first discovered intermittent pulsar, PSR~B1931+24, indicated that the pulsar rotation slows down faster when it is in the ``on'' state than when it is in the ``off'' state~\citep{Kramer06}. This provided quantitative information on the strengths of the magnetosphere electric currents and how particle flows in the pulsar magnetosphere affect the rotation of the pulsar.   To date, noting that there is no clear distinction between long-duration nulling, and short-duration intermittency, only five intermittent pulsars are known~\citep{Lyne17}.

Many mechanisms have been proposed to explain the radio emission (or lack of) in the nulling and intermittent pulsar samples. Several models have been developed that explore how the plasma supply to the magnetosphere affects the global charge distribution (e.g., \citealp{Timokhin10,Li12,Yuen19}). Other models suggest that the radio beam moves out of the line of sight to the pulsar because of changes in the emission geometry (e.g.,  \citealp{Zhang07,Timokhin10}), changes in the properties of the particle acceleration region ~\citep{Szary15}, twists in the magnetic field structure ~\citep{Huang16}, free precession~\citep{Akgun06,Jones12} and/or chaotic ~\citep{Seymour13} and Markov ~\citep{Cordes13} processes.

Long-term observations of intermittent pulsars provide valuable insight in determining the cause of the variability.   In this paper, we present new observations for two intermittent pulsars: PSRs~J1832$+$0029 and J1841$-$0500. 

PSR~J1832+0029 has a period of 533\,ms and was discovered in the Parkes Multibeam Pulsar Survey ~\citep{Lorimer06}. \citet{Lorimer12} showed that it spent around 644\,days in the ``off'' state, it was then ``on'' for 1497 days and then ``off'' again for 835 days.   
The frequency derivatives during ``off'' and ``on'' states were determined to be $-3.2(2)\times 10^{-15}\,{\rm s}^{-2}$ and $-5.44505(7)\times 10^{-15}\,{\rm s}^{-2}$, respectively. 

PSR J1841$-$0500 has a period of 913\,ms and was originally discovered in a search for radio pulsations from the magnetar 1E\,1841$-$045 ~\citep{Camilo12}. The emission for this pulsar remains ``off'' for $\sim$580 days. The corresponding frequency derivative changes from $-1.67\times 10^{-14}\,{\rm s}^{-2}$ to $-4.165(1)\times 10^{-14}\,{\rm s}^{-2}$~\citep{Camilo12} between the ``off'' and  ``on'' states.

This paper is organised as follows.
In \S2, we describe the details of our new observations. Our results are presented in \S3. In \S4, we summarise and discuss these results.

\section{OBSERVATIONS AND DATA PROCESSING}

\begin{table*}  

\centering

\caption{Observations for the two pulsars.} 
\scriptsize
\label{obs}
\begin{tabular}{ccccccc}
\hline
Pulsar name&Receiver& Freq (MHz)& Bandwidth (MHz)& Backend & Obs. Mode & Nobs 
\\
\hline

J1832$+$0029&1050CM &720/728&200&CASPSR &FOLD&3
\\
&H-OH  &1382&400&CASPSR&FOLD&1
\\
&MULTI &1369/1374/1382&256/288/400&AFB/APSR/CASPSR/PDFB2/PDFB3&FOLD&21
\\
&1050CM &732/3094&64/1024&PDFB3/PDFB4&SEARCH&17
\\
&H-OH &1369/1382&256/400&PDFB4/CASPSR&SEARCH&1
\\
&MULTI &1369&256&PDFB4&SEARCH&7
\\
\hline
J1841$-$0500&1050CM &3094/3100&1024&PDFB4&FOLD&13
\\
&MULTI &1382&400&CASPSR&FOLD&7
\\
&10/50cm&3094/3100&1024&PDFB3/PDFB4&SEARCH&31
\\
&H-OH&1369&256&PDFB4&SEARCH&4
\\
&MARS &8500&1024&PDFB4&SEARCH&1
\\
&MULTI &1369&256&PDFB4&SEARCH&12
\\
\hline
 \end{tabular}
\end{table*}

We have continued monitoring observations of PSRs~J1841$-$0500 and J1832$+$0029 using the Parkes 64\,m diameter radio telescope.
We also searched the Parkes data archive \citep{Hobbs11} for publically-available data sets for these pulsars.  
Both search mode (in which single pulses can be studied) and fold mode (in which the incoming data stream has been folded at the known period of the pulsar) observations are included in our data sets. 
For PSR~J1832+0029, we have obtained 50 observations  (16.72 hours in total, consisting of 25 search mode and 25 fold mode observations) between 2005 April 28 and 2018  April 2 (a span of $\sim$13 years).
For PSR~J1841$-$0500, a total of 68 observations (15.96 hours in total, 48 search mode and 20 fold mode observations)
have been collected from 2011 September 2 to 2018 April 2 (a total span of $\sim$7 years).

Observations were obtained using the 10/50\,cm dual-band receiver that provided simultaneous data streams centred at 732 and 3100\,MHz, the H-OH receiver centred at either 1369 or 1382\,MHz, the central beam of the multi-beam receiver in the 20\,cm band centred at 1369\,MHz or the MARS receiver centered at 8500\,MHz.  
Various backend instruments have been used for these observations, including the AFB, APSR, CASPSR, PDFB2, PDFB3, PDFB4 and WBCORR (see, e.g., \citealt{Manchester13} for details).  For each pulsar, the receivers, center frequencies, bandwidths,  backends, observation modes and number of observations are listed in Table \ref{obs}.   All these observations are available for download from the Parkes data archive (\url{data.csiro.au}) with the majority of the observations obtained using the P863 project code.

To analyze the search mode observations, we used {\sc dspsr}~\citep{Hotan04} to extract individual pulses, which were further processed using the {\sc psrchive} software suite.  Fold-mode observations were directly analysed using {\sc psrchive}.  Data affected by narrow-band and impulsive radio-frequency interference were flagged and 5 percent of the band-edges were removed.  A switched calibration signal was recorded before or after the observation.  The data were calibrated with associated calibration files using the {\sc psrchive} program {\sc pac} which flattens the bandpass, flux calibrates and transforms the polarzation products to Stokes parameters (see e.g., \citealt{2019RAA....19..103X}). Topocentric times of arrival (ToAs) were obtained by cross correlating the mean pulse profile with a noise-free template using {\sc psrchive}. Timing residuals were formed using the {\sc tempo2} software package ~\citep{Hobbs06}.  The analysis of fluctuation spectra was carried out with the {\sc psrsalsa} package~\citep{Weltevrede16}.

\section{RESULTS}

\subsection{PSR~J1832+0029}

PSR~J1832$+$0029 is relatively weak (the flux density in the 20\,cm observing band is 0.14\,mJy; \citealt{Lorimer06}).  We therefore require approximately at least a 300\,s integration to detect this pulsar (during the ``on'' state) in the 20\,cm band and $\sim$600\,s with the narrower-band 50\,cm receiver.  We have therefore removed from further analysis any observations which had shorter integration times or were significantly affected by radio frequency interference.    

\begin{figure}
\centering
\includegraphics[width=85mm]{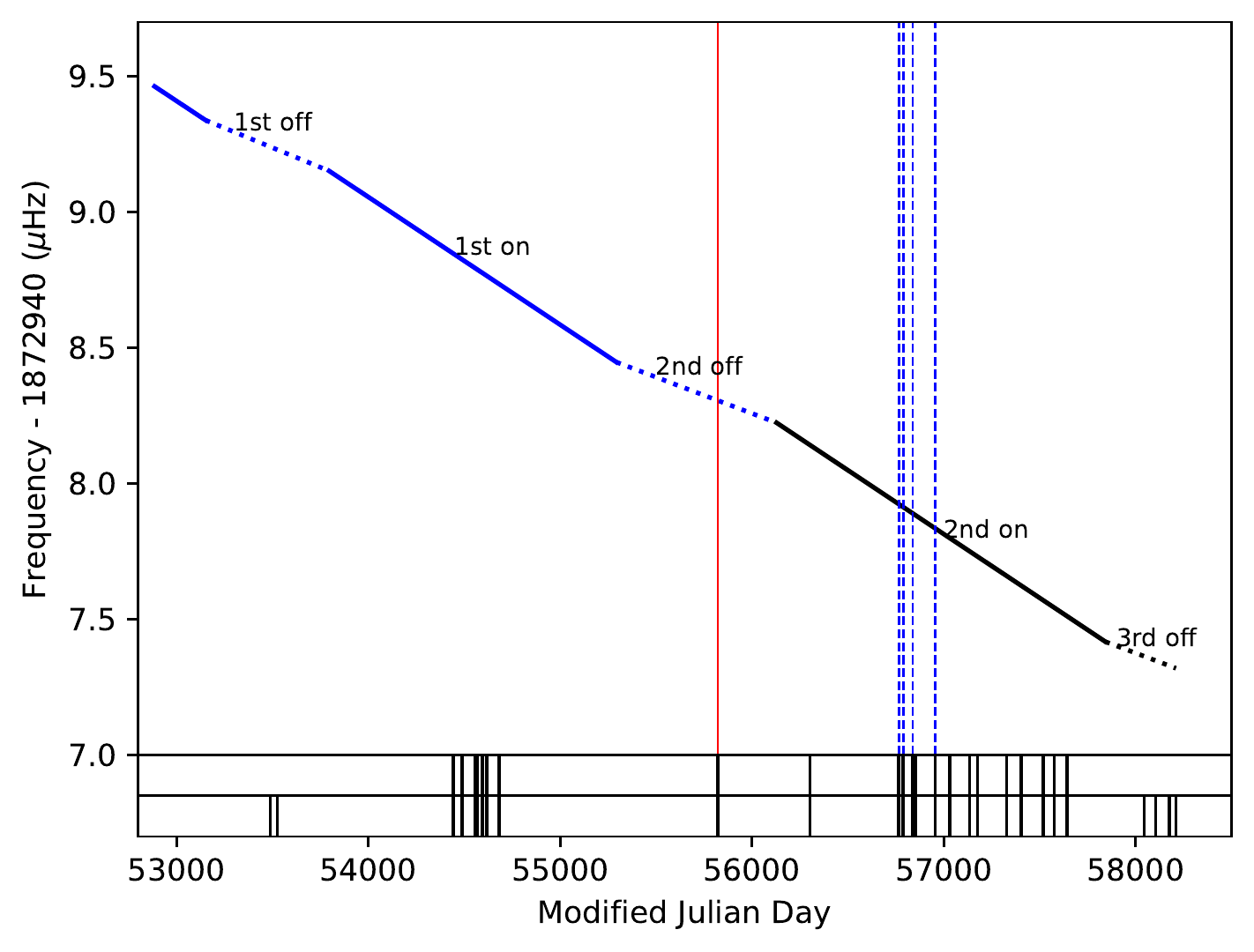}
\caption{Upper panel: pulse frequency versus MJD for PSR~J1832+0029. The solid lines represent the pulse frequencies determined phase connected timing solutions. The dotted lines are the extrapolated pulse frequency when the pulsar was not detected.  The data points for the blue, solid lines were obtained from~\cite{Lorimer12}. The vertical lines are described in the text. The lower panel is split into two.  In the top the MJDs of the 37 observations in which the pulsar was detected are shown. The bottom shows the MJDs for all of our observations of this pulsar.}\label{18321}
\end{figure}

This pulsar was detected in 37 of the 50 remaining observations. 17 of these were recorded in search mode and the remaining 20 were fold mode observations.  The observation and detection epochs are shown in the bottom and middle panels of Figure~\ref{18321}, respectively. The rotational frequency history is shown in the upper panel, the blue lines correspond to data in ~\citet{Lorimer12} and the black lines are from our new observations.  We have  measurements spanning one new ``on'' state (the 2nd on in Figure~\ref{18321}) and one new ``off'' state (the 3rd off in Figure~\ref{18321}). 

We have determined the timing solutions for the pulsar during the new ``on'' state (Table~\ref{1832t2}). 
The duration of the new “on” state is $1714\pm200$ days and the corresponding frequency derivatives are $-5.446(1)\times10^{-15}\, {\rm s}^{-2}$, which is consistent with the result of~\citet{Lorimer12} of $-5.44505(7)\times 10^{-15}\,{\rm s}^{-2}$. The frequency derivatives during different ``on'' states are consistent. However, the emission of this pulsar is not stable during ``on'' state.  In four of our six simultaneous observations at 10\,cm and 50\,cm bands, the pulsar was clearly detected in the 50\,cm band, but not in the 10\,cm band. However, we note that this pulsar is always weak in the 10\,cm band. We have indicated these observations using vertical, dashed lines in Figure~\ref{18321}.

\begin{figure}
  \centering
  \includegraphics[width=80mm]{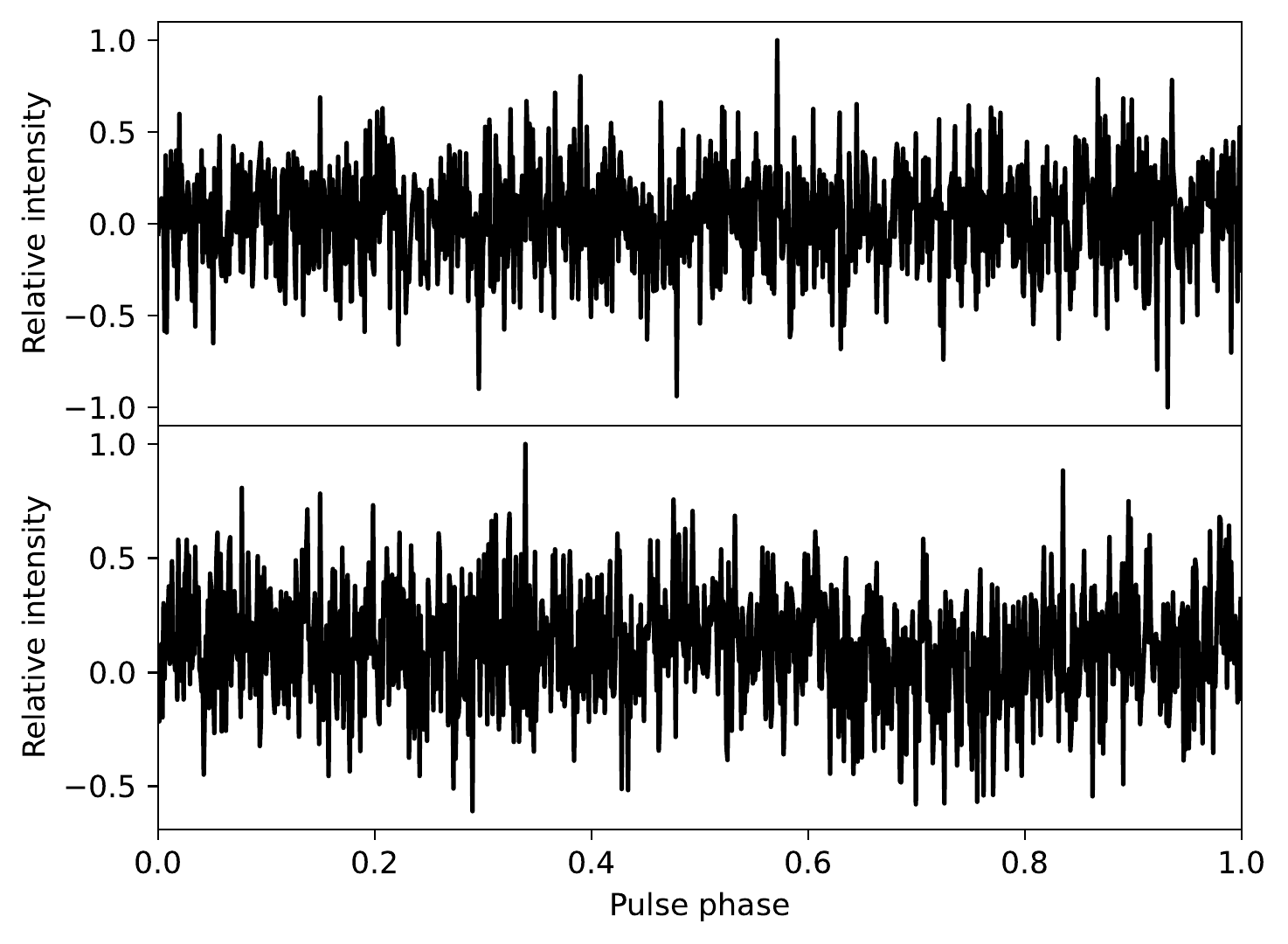}
  \caption{The accumulated pulsar profiles during ``off'' states for PSRs J1832+0029 (upper panel) and J1841$-$0500 (lower panel), respectively. All the profiles are normalised.}\label{prof}
\end{figure}

The pulsar currently remains ``off'' in its third ``off'' state (after MJD 57842).  Our last processed observation (in which the pulsar remained undetectable) was on the 2018 April 2 (MJD 58210) and so the pulsar has not been detected for at least 368 days. We have summed the folded-profiles for the observations during this ``off'' state for observations recorded using the PDFB4 signal processor centred at 1369 MHz. The accumulated profile is shown in upper panel of Figure~\ref{prof} and no significant emission is detected and the flux density limit is  $\sim$0.2\,mJy.

\begin{figure}
\centering
\includegraphics[width=80mm]{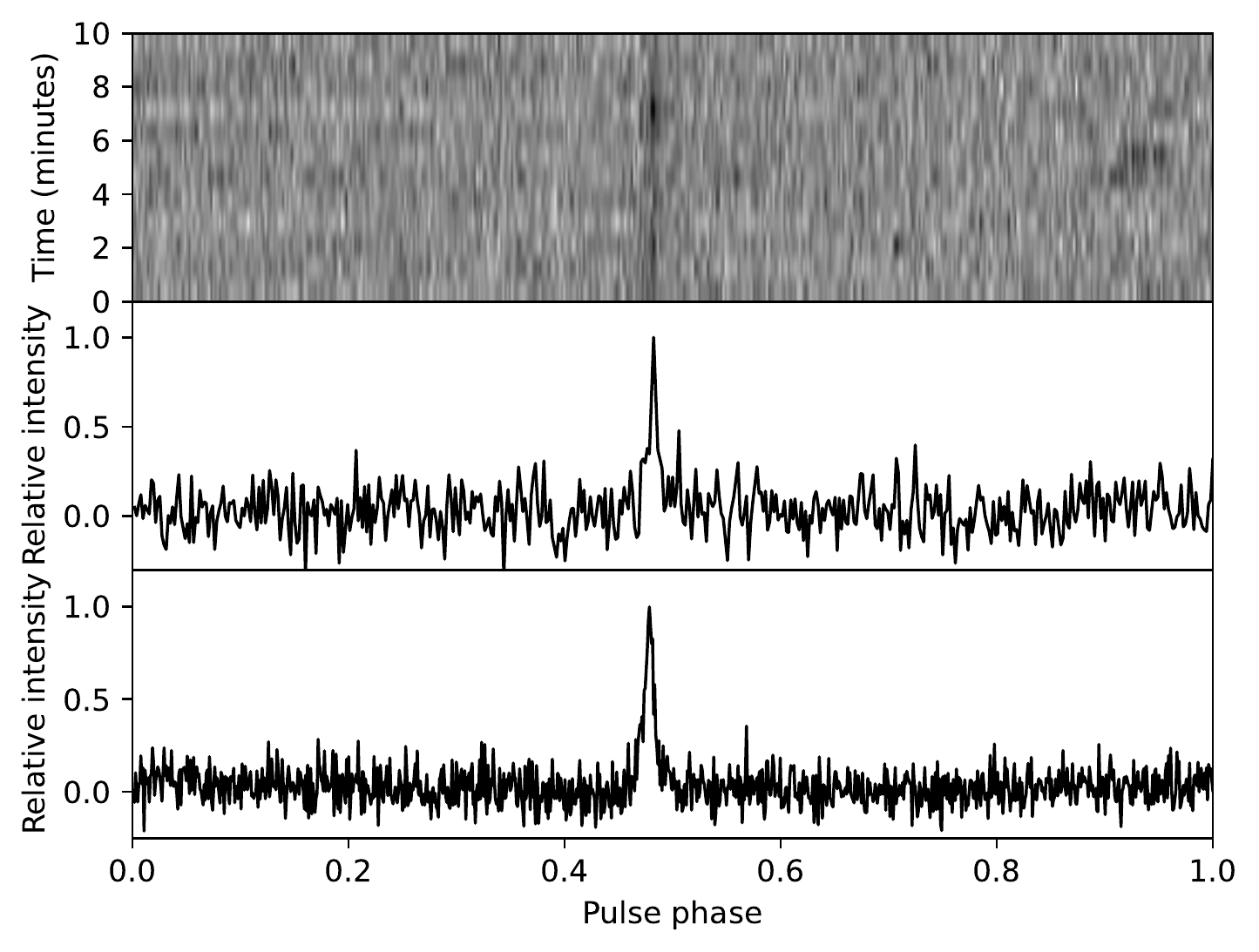}
\caption{The pulse intensity as a function of pulse phase and time (upper panel) and its averaged pulse profile (middle panel). We note that this emission occurred during the second ``off'' state of PSR J1832+0029 on 2011 September 17 observation. A typical accumulated profile of the pulsar during ``on'' state is shown in lower panel. Both profiles are normalised. 
}\label{1832weak}
\end{figure}

We note that we only have a single Parkes observation during the second ``off'' state and in our observation the pulsar is ``on''. The duration of this observation, on 2011 September 17, is only 10 minutes and the integrated pulse profile is weak, but is clearly detectable, has the correct dispersion measure and has a similar profile compared with ``on'' state (see Figure~\ref{1832weak}).  This ``off'' state was defined by \citet{Lorimer12} using multiple observations from the Lovell radio telescope and two very sensitive observations using the Arecibo telescope (placing an upper bound on radio emission in the 20\,cm band of 1.6$\mu$Jy.    Our observation suggests that the pulsar ``flickers on'' for very short durations during the ``off'' state.  This is similar to PSR~J1841$-$0500 which is described in more detail below, where \citet{Camilo12} reported that the pulsar turned off in brief episodes during an ``on'' state.

\begin{table} 
\caption{Parameters for PSR J1832+0029}
\scriptsize
\label{1832t2}
\centering
\begin{tabular}{cc}

\hline
 parameters    &       value\\
\hline
Data span (MJD) & 53152$-$53796 (1st ``off'' phase)
\\
Frequency derivative ($10^{-15}\, {\rm s}^{-2}$) & $-3.2(2)$
\\
Duration time (day) & $644$\\
\hline
Data span (MJD) & 53796$-$55293 (1st ``on'' phase)
\\
Frequency derivative ($10^{-15}\, {\rm s}^{-2}$) & $-5.44505(7)$
\\
Duration time (day) & $1497$\\
\hline
Data span (MJD) & 55293 $-$56128 (2nd ``off'' phase)
\\
Frequency derivative ($10^{-15}\, {\rm s}^{-2}$) & $-3.08(5)$
\\
Duration time (day) & 835\\
\hline
Data span (MJD) & 56128$-$57842 (2nd ``on'' phase)
\\
Spin frequency (Hz) & 1.87294771837(1)
\\
Frequency derivative ($10^{-15}\, {\rm s}^{-2}$) & $-5.446(1)$
\\
Reference epoch (MJD) & 57203\\
Duration time (day) & $1714\pm200$\\
\hline
Data span (MJD) & 57842$-$58210 (3rd ``off'' phase)\\
Duration time (day) & 368\\
\hline
 \end{tabular}
\end{table}

\subsection{PSR~J1841$-$0500}

\begin{figure}
\centering
\includegraphics[width=85mm]{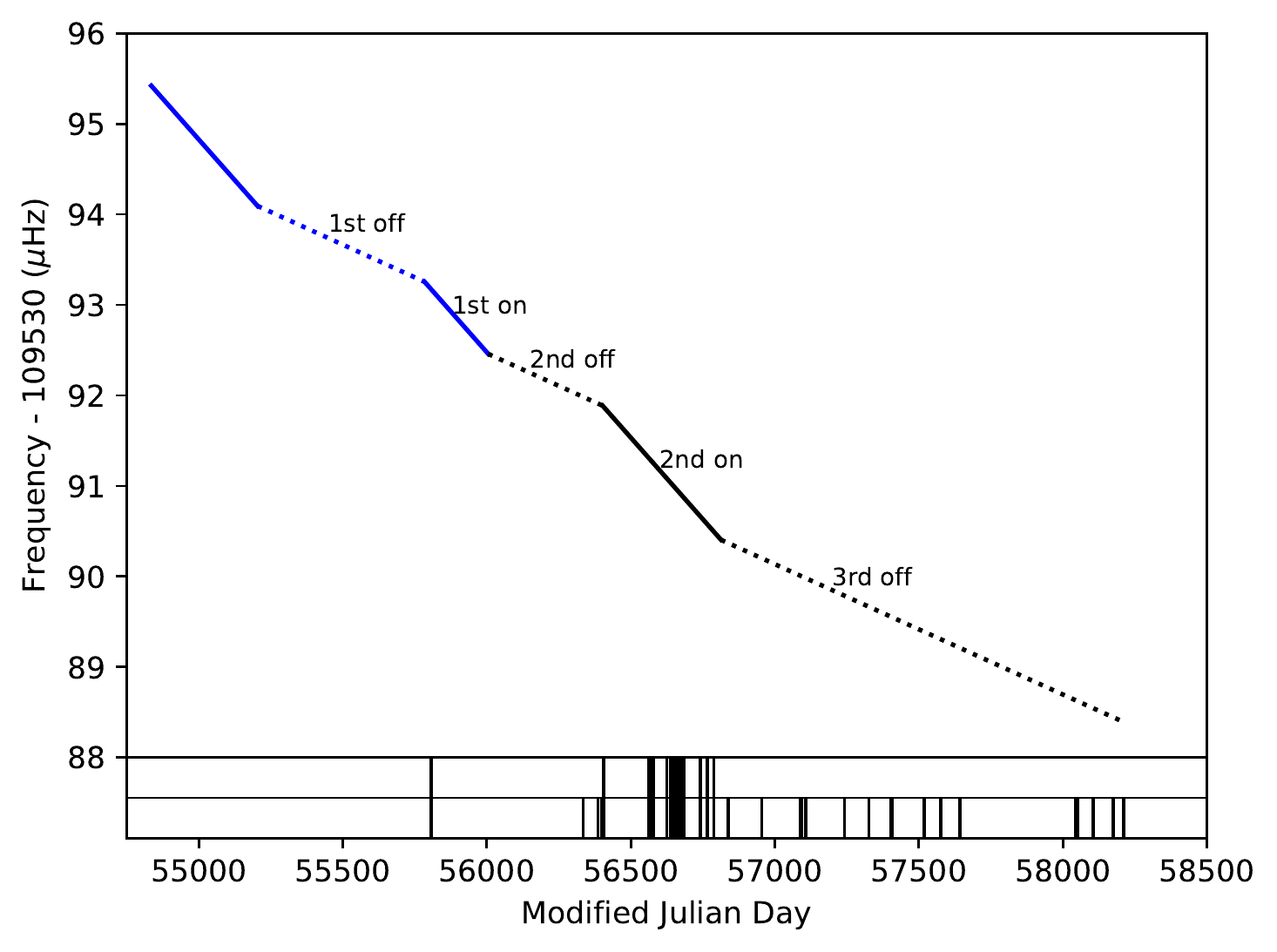}
\caption{The pulse frequency versus MJD for J1841$-$0500. The panels are the same as in Figure 1. The data points for the blue lines are from ~\citealt{Camilo12}.
}\label{18411}
\end{figure}

Radio emission from PSR~J1841$-$0500 was detected in 32 out of our 68 observations. These detections were made in 22 search mode data files and 11 fold mode files with the observing frequency centred at 3.1\,GHz. In Figure~\ref{18411}, we show our observation epochs in the bottom panel and the epochs when the pulsar was detected in the middle panel. The upper panel shows the rotational frequency as a function of epoch, in which the data shown in blue are from~\citet{Camilo12} and in black from our observations, which cover one new ``on'' state (the 2nd on in Figure~\ref{18411}) and two new ``off'' states (the 2nd and 3rd off in Figure~\ref{18411}). 

We determined the timing solution for the pulsar in the ``on'' state from our observations. The result is listed in Table \ref{1841t2}. 
The duration of the new ``on'' state is $411\pm27$ days and the frequency derivative is $-4.173(1)\times10^{-14}\, {\rm s}^{-2}$, which is consistent with the result of~\citet{Camilo12} of $-4.165(1)\times 10^{-14}\,{\rm s}^{-2}$.
The duration of the second ``off'' state cannot be determined accurately because we do not have a large number of observations during that time. Assuming the frequency derivative of this state is as same as the first ``off'' state, then the duration time can be extrapolated to 397\,days, which is shorter than that of the first ``off'' state (which was 580\,days).
During the third ``off'' state (after MJD 56813), the radio emission has remained undetected for at least 1397 days in our observations during this time.  There is also an interval of around 400 days (MJD 57642 to 58043) during which we have no observations.
It is possible that pulsar switched to the ``on'' state during this span and, if so, the duration of the third ``off'' state for PSR~J1841$-$0500 would reduce to 829 days, which is still longer than the former two ``off'' states.

In contrast to ~\citet{Camilo12}, we did not detect any short timescale on-off switches.  \citet{Camilo12} showed that the pulse may switch off during ``on'' states for up to a few percent of the available time.  Our observations are consistent with the same statistics, but we made no detection of a short-scale switch.

We have used all of the available observations during the ``off'' state to search for any detectable emission. The lower panel of Figure~\ref{prof} presents the accumulated profile from summing all the observations obtained using the PDFB4 signal processor and recorded with a central observing frequency of 1369\,MHz. No significant emission has been found and the flux limit is 0.11\,mJy.     

\begin{figure}
  \centering
  \includegraphics[width=80mm]{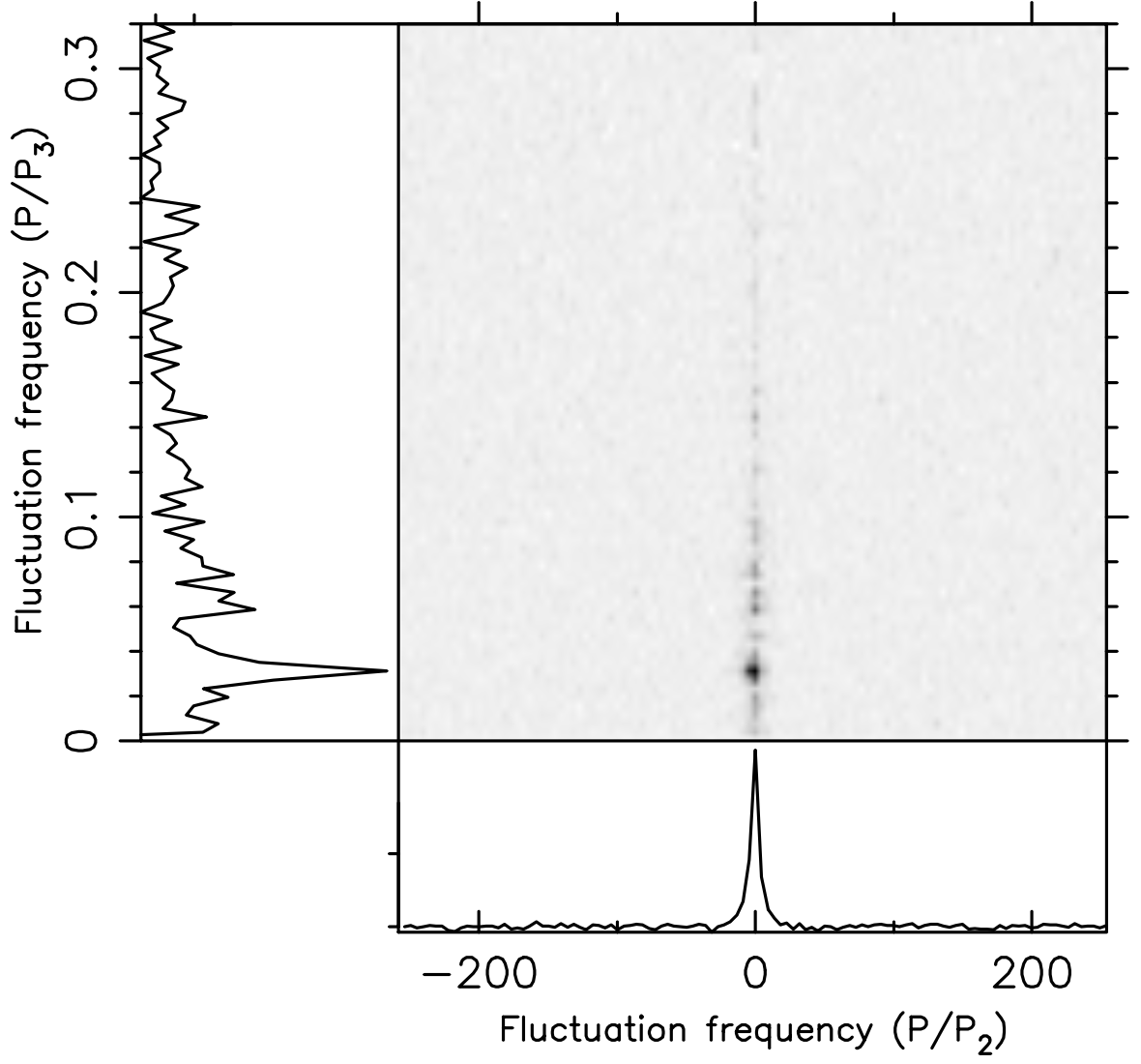}
  \caption{The 2DFS of PSR J1841$-$0500. The power in the 2DFS is vertically integrated, producing the bottom plots and it is horizontally integrated, producing the side-panels of the spectra. }\label{18412dfs}
\end{figure}

\begin{figure}
  \centering
  \includegraphics[width=80mm]{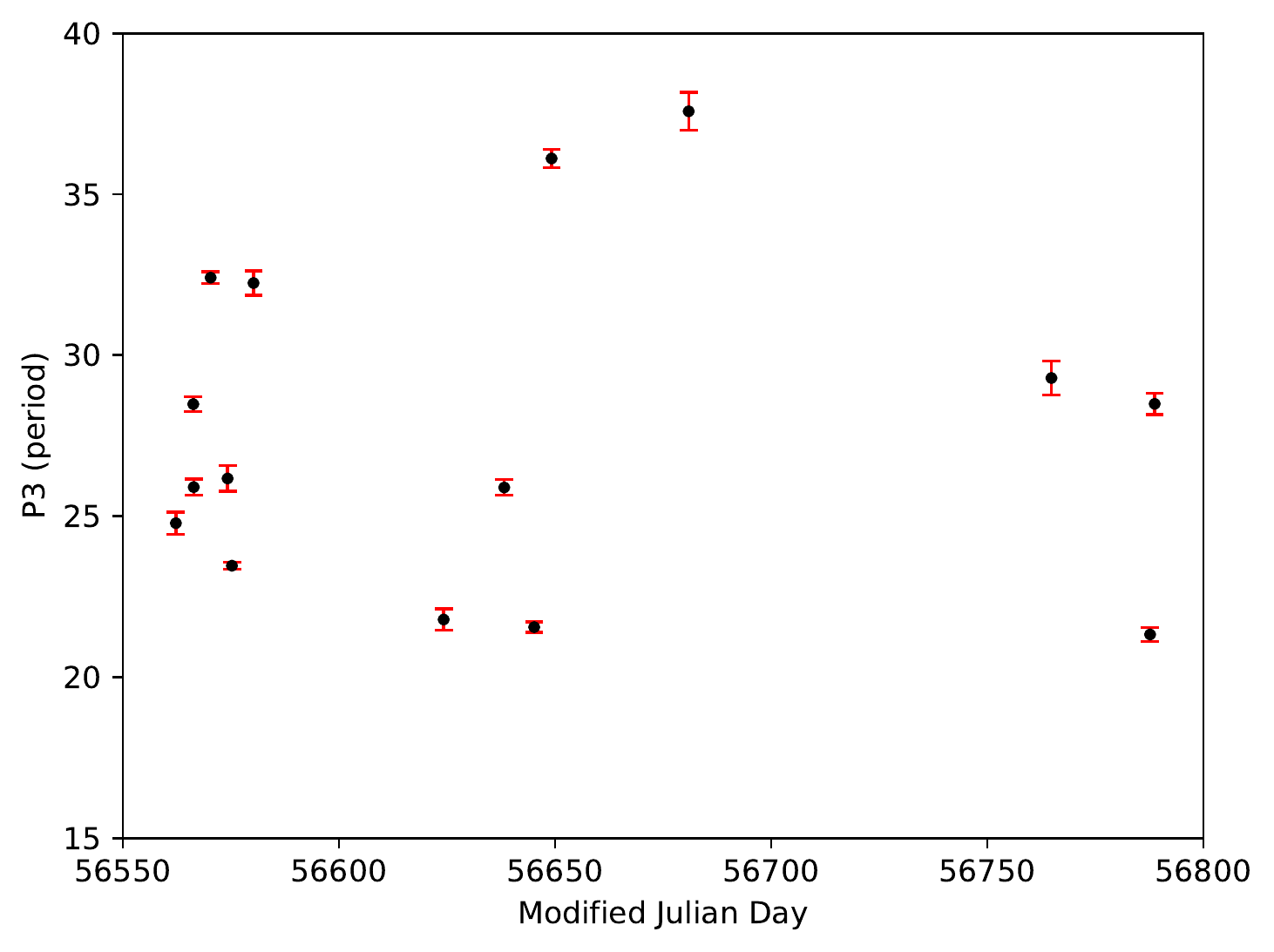}
  \caption{$P_3$ versus MJD for PSR J1841$-$0500. The $P_3$ values (black dots) are obtained using the 2DFS described in the text and the error bars are based on the root-mean-square of the noise.}\label{p3}
\end{figure}

In the ``on'' state we can detect many of the individual pulses from the pulsar. We noticed an intensity modulation in these single pulse intensities and applied a two-dimensional fluctuation spectrum (2DFS, \citealt{Edwards02}) to explore this further.    The 2DFS is used for analysing sub-pulse modulation with the periodicity of intensity modulation $P_{3}$ and the characteristic horizontal time separation between modulation bands $P_{2}$.
We obtained that $P_{2}=0$ and $P_{3}=32.40\pm0.18\,P$ where $P$ is the pulse period (Figure~\ref{18412dfs}). This suggests that this pulsar has a significant periodic pulse intensity modulation, but as $P_2 = 0$ this is not caused by traditional sub-pulse drifting.  We note that this flux density modulation is not constant for all of our observations. Figure~\ref{p3} presents the single pulse intensity modulation periods measured for all of the detections; the values range from 21 to 36\,$P$.

PSR~J1841$-$0500 has a high DM (about 532\,${\rm cm^{-3} pc}$) and therefore the observed emission is affected by interstellar scattering~\citep{Scheuer68}, giving rise to an exponential decay of the pulse with the scatter broadening time $\tau_{\rm sc}$~\citep{Williamson72}.  Our observations allow us to provide the first measurement of $\tau_{\rm sc}$ for this pulsar in the 3.1\,GHz band. We obtained  $\tau_{\rm sc}=28.39\pm0.27\,{\rm ms}$.

\begin{figure}
 \centering
\includegraphics[width=80mm]{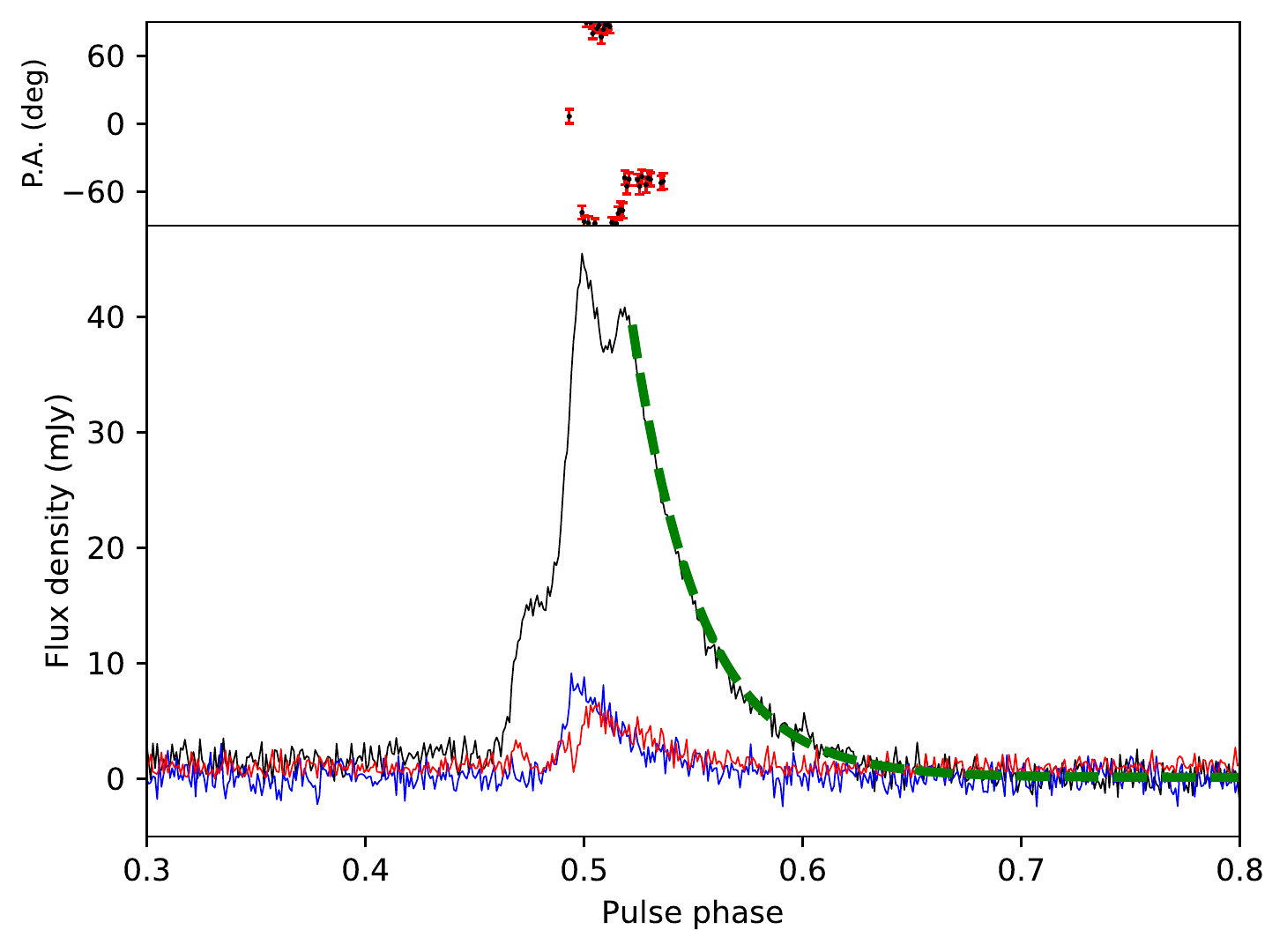}
\caption{Polarization profile at 3.1 GHz for PSR J1841$-$0500. The black line shows total intensity, red linear polarization and blue circular polarization, respectively. The green dashed line is the fit to the tail of the profile.
The position angles (black dots) with the error bars (red bars) of the linearly polarised emission are shown in the upper panel.}\label{1841p}
  \end{figure}

We are also able to present the first polarization profile (at 3.1\,GHz) for this pulsar.  The polarization (see Figure~\ref{1841p}) is similar to that of 2\,GHz profile (see~\citealt{Camilo12}). The linear polarization is stronger than the circular polarization. The mean flux density at 3.1\,GHz determined from this observation is $2.59\pm0.01$\,mJy.  Combining the mean flux density at 2, 5 and 9 GHz (see~\citealt{Camilo12} ),  we obtained a spectral index of $-2.14\pm0.17$.
We also determined the rotation measure (RM) to be $-3058\pm 3\,{\rm rad\,m^{-2}}$, which is consistent with the value of -3000 presented by ~\citet{Camilo12}.

\begin{table} 
\caption{Parameters for PSR J1841$-$0500}
\scriptsize
\label{1841t2}
\begin{tabular}{cc}
\hline
 parameters    &       value\\
\hline
Data span (MJD) & 55204$-$55784 (1st ``off'' phase)
\\
Frequency derivative ($10^{-14}\, {\rm s}^{-2}$) & $-$1.67
\\
Duration time (day) & $580$
\\
\hline
Data span (MJD) & 55784$-$56005 (1st ``on'' phase)
\\
Frequency derivative ($10^{-14}\, {\rm s}^{-2}$) & $-$4.17(1)
\\

Duration time (day) & 221\\
\hline
Data span (MJD) & 56005$-$56402 (2nd ``off'' phase)
\\
Frequency derivative ($10^{-14}\, {\rm s}^{-2}$) & $-$1.67
\\
Duration time (day) & $397$
\\
\hline
Data span (MJD) & 56402$-$56813 (2nd ``on'' phase)
\\
Spin frequency (Hz) & 1.09539127017(5)
\\
Frequency derivative ($10^{-14}\, {\rm s}^{-2}$) & $-$4.173(1)
\\
Reference epoch (MJD) & 56573\\
Duration time (day) & $411\pm27$\\
\hline
Data span (MJD) & 56813$-$58210 (3rd ``off'' phase)\\
Duration time (day) & 1397\\
\hline
 \end{tabular}
\end{table}

\section{DISCUSSION AND CONCLUSIONS}

We have presented new observations of two intermittent pulsars, PSRs~J1832+0029 and J1841$-$0500.
In general the results are consistent with previous observations. Both pulsars continue to switch on-and-off on timescales much longer than the rotational period of the pulsars.
We have noted clear emission during a single observation during the ``off'' state for PSR~J1832$+$0029. 
Similarly, \citet{Camilo12} reported a short duration emission cease during the ``on'' state for PSR J1841$-$0500.
Together these results suggest that the emission state can switch rapidly and the long ``on'' and ``off'' durations detected for these intermittent may simply be because of the poor observing cadence.  More regular observing (which will arise naturally from telescopes with wide-fields-of-view such as the Molonglo observing program, \citealt{Bailes17}) will enable us to clarify the switching rate for these pulsars.
%Similarly, the emission for some nulling pulsars does not disappear completely during the nulling state, such as PSRs J0941$-$39~\citep{Burke12} and B0826$-$34~\citep{Esamdin12}.
%However, we cannot confirm whether they have the similar physical mechanism. 

The quasi-periodic oscillations detected in the flux density for PSR~J1841$-$0500 are also unexpected. Such oscillations are seen in pulsars that exhibit sub-pulse drifting, but we note that $P_2 = 0$ in our analysis suggesting that drift-bands are not present.  
\citet{Basu16} found similar behaviour in a sample of pulsars (that did not include an intermittent pulsar) and noted that the flux density variation was associated with the core component in the pulse profiles. For these pulsars, the rotation energies were all larger than $5\times10^{32}\,{\rm erg\,s^{-1}}$. The spin down energy loss rate for PSR~J1841$-$0500 is $1.8\times10^{33}\,{\rm erg\,s^{-1}}$~\citep{Camilo12}, which supports the theory that there may be a link between this pulsar and those studied by \citet{Basu16}. 

The physical process that underlies the intermittent emission behaviour is still not clear.  In some cases, a clear periodicity (or quasi periodicity) has been detected in the switching time-scale. For PSR~1931+24, \citet{Cordes08} theorized that the quasi-periodic intermittent emission behaviour is driven by a debris disk around the pulsar, with a 40 day orbit. The interaction between the disk and the pulsar magnetosphere  can periodically halt the radio emission from the pulsar. Similarly, \citet{Rea08} proposed that the pulsar accretion from a low mass companion in the orbital periastron could account for the quasi-periodic radio emission. However, no X-ray emission, which is predicted by both of these two scenarios, has been detected for the intermittent pulsar sample.  However, the on-off cycle timescales for both of the pulsars described in this paper, PSR~J1832+0029 and PSR~J1841$-$0500, are not quasi-periodic.  Similarly, \citet{Lyne17} reported a significant evolution of on-off cycle timescales of an intermittent pulsar J1929+1357, which cannot result from a simple pulsar accretion process.

High cadence observing over long data spans, with wide receiving bandwidths and highly sensitive telescopes is necessary to produce more stringent constraints on the theoretical models.   Our Parkes observations are not sensitive enough to detect high signal-to-noise single pulses from most of these intermittent pulsars, but we note that many of these pulsars lie within the MeerKAT and Five-hundred-metre Spherical Radio Telescope (FAST) sky coverage. These, more sensitive, telescopes will enable us to provide more stringent constraints on emission during the ``off'' states as well as allowing us to study whether the flux density oscillations on a single pulse level are related to the emission state switching.  The Parkes ultra-wide-bandwidth receiver \citep{Hobbs19} will also enable these pulsars to be studied at higher observing frequencies than currently available. 

%The intermittency behaviour of the intermittent pulsar may be not associated with the pulsar accretion process.
%The sample of the intermittent pulsar are still rare. 
%More observations are needed to restraint the theory models.

\section*{Acknowledgments}
The Parkes radio telescope is part of the Australia Telescope National Facility which is funded by the Commonwealth of Australia for operation as a National Facility managed by CSIRO. This paper includes archived data obtained through the CSIRO Data Access Portal (http://data.csiro.au). This work is supported by the Youth Innovation Promotion Association of Chinese Academy of Sciences, Heaven Lake Hundred-Talent Program of Xinjiang Uygur Autonomous Region of China, the National Key Research and Development Program of China (No.2016YFA0400804 and No.2017YFA0402602), the Strategic Priority Research Program (B) of the Chinese Academy of Sciences (No.XDB230102000), CAS-MPG LEGACY funding and the Operation, Maintenance and Upgrading Fund for Astronomical Telescopes and Facility Instruments, budgeted from the Ministry of Finance of China (MOF) and administrated by the Chinese Academy of Sciences (CAS). R.M.S. acknowledges  Australian Research Council grants FT190100155 and CE170100004.

\end{document}